\documentstyle[11pt,aaspp4]{article}

\begin{document}

\title{The Effect of Self-gravity of Gas on Gas Fueling in Barred Galaxies with a Supermassive Black Hole}

\author{Hiroyuki Fukuda and Asao Habe}
\affil{Division of Physics, Graduate School of Science, Hokkaido University, Sapporo 060-0810, Japan}
\author{Keiichi Wada\altaffilmark{1}}
\affil{National Astronomical Observatory, Mitaka, Tokyo 181-8588, Japan}
\altaffiltext{1}{Johns Hopkins University, Baltimore, MD, 21218, USA}

\begin{abstract}
In our previous paper, we have shown that a gas disk in the nuclear region of
a barred galaxy which contains a central supermassive black hole (SMBH)
rapidly evolves into a nuclear gas ring by the effect of an additional inner
Lindblad resonance caused by the SMBH.
In this paper, we investigate the fate of the gas ring, involving self-gravity
of gas, using two-dimensional hydrodynamical simulations.
We find that the gas ring becomes gravitationally unstable for a gas surface
density of gas above a critical value, 
and fragments into several gas clumps.
Some denser clumps increase their mass via the 
accretion of the surrounding gas and collisions with other
clumps, and finally a very massive gas clump ($M\sim 10^7 M_\odot$) is formed.
Due to the torque from the most massive clump, 
a part of the gas in the ring loses its angular momentum and falls into
the galactic center. As a result, a nuclear gas disk ($R\sim 50$ pc) is formed 
around the SMBH. 
The accretion rate for $R<50$ pc attains about $1 M_\odot$ yr$^{-1}$ for
3.5$\times 10^7$ yr. 
At the final phase of the bar-driven fueling,
self-gravity is crucial for the angular momentum transfer of the gas.
This is a new mechanism for gas fueling to the vicinity of the SMBH.
\end{abstract}


\section{Introduction}
Gas fueling is supposed to trigger and support the nuclear activities in
galaxies, such as nuclear starbursts or AGNs (Shlosman, Begelman \& Frank 1990).
Many observational, theoretical and numerical studies have suggested
that a non-axisymmetric potential due to a stellar bar or the galaxy-galaxy
interactions can induce gas fueling for the galactic central region
(Noguchi 1988; Shlosman, Begelman \& Frank 1990; Barnes \& Hernquist 1992;
Knapen et al. 1995; Benedict et al. 1996).
Numerical simulations of gas in a bar-like potential 
have shown that the gas forms spirals or oval ring-like structure near
 the inner Lindblad resonances (ILRs),
and the gas is accumulated into an oval ring-like structure when the ILR exists
(Elmegreen 1994 and references therein).
However, since the ILR is usually located at several 100 pc or a few kpc from
the galactic center, the angular momentum transport due to the bar itself is
not enough to fuel the gas in the vicinity of the AGN.
\footnote[2]{
If the bar
is very strong (e.g. the axial ratio $a/b \ga 3$),
the gas can fall toward the center (Athanassoula 1992).
In Athanassoula's strong bar models,
she used a different definition for {\it ILR}: the model has one {\it ILR}
if it has an $x_2$ orbit family extending all the way to the center.
Therefore, the strong bar model, in which all gaseous orbits are $x_1$ like
and as a result there is a strong radial flow of the gas, does {\it not}
have an {\it ILR} in Athanassoula's definition. One should note that
this is different from the analytic definition of the ILR 
(i.e. $m(\Omega_0-\Omega_b)=\pm \kappa_0$), which is derived from 
a linear perturbation theory in a weak bar-like potential.
 It has been often referred like that {\it if there is
no ILR, the gaseous inflow can reach to the center}. However, this is not
true for a weak bar (e.g. Wada and Habe 1992).
}

Recent observations suggest that many galaxies have central supermassive 
black holes (SMBHs) (Ford et al.
1994; Harms et al. 1994; Miyoshi et al. 1995; Kormendy \& Richstone 1995).
In a previous paper (Fukuda, Wada \& Habe 1998, hereafter Paper I), 
we have pointed out that a SMBH in the central
region of a weak bar may change the gas dynamics within 1 kpc from the center
drastically.
There, we have investigated the dynamics of a non-self-gravitating gas disk
near the additional ILR (hereafter we call it a nuclear Lindblad resonance
(NLR)) which is caused by the SMBH inside the usual ILRs, using the smoothed
particle hydrodynamics (SPH) method.
We have found that strong trailing spiral shocks
are formed in the NLR region even if the bar is weak,
 and the gas in the vicinity of the shock is
rapidly fueled into the central region of the galaxy.
As a result, a large amount of gas concentrates into the nuclear ring in a
dynamical time scale of the central region.
We have deduced the critical surface density of the initial gas disk
for which the gas ring becomes gravitationally unstable.
The critical surface density of the initial gas disk is of the same order
as the surface density of gas in normal disk galaxies.
Therefore, the nuclear gas ring is expected to be gravitationally
unstable, and the further evolution of the ring must be affected by the
self-gravity of gas. This is the motivation for our study presented here.

It has been shown that the self-gravitating gas which is
concentrated near the usual ILRs could be fueled to
much closer towards the galaxy center
(Fukunaga \& Tosa 1991; Wada \& Habe 1992, 1995).
Heller \& Shlosman (1994) studied the evolution of the gas distribution in
a globally unstable (i.e. bar forming) stellar disk using self-consistent
simulations including star formation.
They placed a `seed' SMBH at the center and investigated the accretion
onto the SMBH.
However, the role that the NLR caused by the SMBH plays 
in the fueling process is not clear
in their study.

In this paper, we investigate the further evolution of the
gas ring which is formed by the effect of the NLR using the SPH method
taking into account the self-gravity of the gas.

In section 2, we describe the numerical method and the model of a galaxy with
an SMBH and a gas disk.
Numerical results are presented in section 3.
In section 4, we summarize our results, and discuss the effects of the
self-gravity of gas and the observations related to the numerical results.

\section{Numerical Methods and Models}

\subsection{Numerical method}

The numerical method and models are based on Paper I.
The hydrodynamical part of the method is the same as that in Paper I,
but with more numerical resolution.
We use the SPH method (Lucy 1977; Gingold \& Monaghan 1982) to solve the
hydrodynamical equations.
In the SPH method, physical quantities at a spatial point are represented by
the sum of the contributions from the neighboring SPH particles at that point.
The contribution from an SPH particle is expressed by a smoothing kernel.
We choose the Gaussian smoothing kernel and take the symmetrized formula in the
pressure gradient and artificial viscosity to treat shock phenomena in a gas
disk correctly, as in Paper I.
More details of our SPH code is given in Paper I.

We use the hierarchical tree method (Barnes \& Hut 1986) for the calculation
of the self-gravity of gas and the neighbor search in the SPH code.
The tolerance parameter $\theta=0.5$, and the quadrupole moment expansion
for the self-gravity calculation is used in the tree method.
The gravitational softening length of the SPH particles is set equal to the
smoothing length, $\epsilon = h$.
To avoid excessive computing time, we impose a lower limit,
$h_{\rm min} = 0.001$, on the smoothing length.
This lower limit is small enough to resolve nuclear gas structure shown in
section 3.

\subsection{Model galaxy}

We assume that the model galaxy is composed of a stellar disk, a bulge, a 
stellar bar, and an SMBH at the center of the galaxy.
The stellar bar is assumed to be `weak', i.e. the 
non-axisymmetric component of the potential is less than 10 percent of
the axisymmetric one.
This is because 1) we investigate the gas dynamics in the central few
kpc region where the barred component is weak, and 2) a strong stellar bar
could be destroyed due to the SMBH (Hasan \& Norman 1990;
Hasan, Pfenniger \& Norman 1993; Norman, Sellwood \& Hasan 1996;
Hozumi \& Hernquist 1998).
The weak stellar bar is treated as an external fixed potential. We do not
consider a dark halo component, because we are interested in gas dynamics in
the inner region of the galaxy.

The axisymmetric potential of the galaxy is assumed to be 
the Toomre disk (Toomre 1963) :
\begin{equation}
\Phi_{\rm axi}(R)=-\frac{GM_{\rm axi}}{(R^2+a^2)^{1/2}},
\end{equation}
where $M_{\rm axi}$ is the dynamical mass of the disk and the bulge, and $a$ is a 
core radius which roughly corresponds to the size of the bar.
We assume the barred potential to be :
\begin{equation}
\Phi_{\rm bar}(R,\theta)=\epsilon(R)\Phi_{\rm axi}\cos2\theta,
\end{equation}
where $\epsilon(R)$ is given by
\begin{equation}
\epsilon(R)=\epsilon_0\frac{aR^2}{(R^2+a^2)^{3/2}},
\end{equation}
and $\epsilon_0$ is a parameter which represents the strength of the bar to the
axisymmetric component.
We assume the potential of the SMBH :
\begin{equation}
\Phi_{\rm BH}(R)=-f_{\rm BH}\frac{GM_{\rm axi}}{(R^2+{a_{\rm BH}}^2)^{1/2}},
\end{equation}
where $f_{\rm BH}$ is a mass ratio of the SMBH to $M_{\rm axi}$, and $a_{\rm
BH}$ is a softening parameter.

Unit mass, length, and time are $3{\times}10^{10}$ $M_\odot$, 5 kpc,
and $1.75\times10^8$ yr, respectively. We choose $a = 0.4$, $M_{\rm axi}
=1.63$, $\epsilon_0= 0.1$, and $a_{\rm BH} = 0.001$.
For the pattern speed of the bar and mass of the SMBH, 
we use model B in Paper I, in which $f_{\rm BH}=0.01$, and $\Omega_{\rm bar}
=3.5$. In this model, the radius of the NLR is $0.14$, and the inner ILR is
located at $R=0.37$.
The gravitational potential of the SMBH is dominant within $R\sim0.05$.

\subsection{Gas disk models}

We distribute $5\times10^4$ SPH particles randomly within $R=0.7$ in order to
represent a uniform density disk at $t=0$.
The initial rotational velocity is given in order to balance the centrifugal
force caused by the axisymmetric potential, since the distorted part of the
potential is very weak.
The rotational period at $R=0.7$ is about 0.7.
We assume that the gas is isothermal and that its temperature is $10^4$ K,
which corresponds to a sound speed of $C_{\rm s} \sim 10$ km s$^{-1}$.

In Paper I, we estimated the self-gravity of the finally formed gas ring
in order to investigate whether the gas ring is gravitationally unstable or not.
We used the Toomre's $Q$ value as an indicator of self-gravitational
instability.
The Toomre's $Q$ value of the gas ring in model B is
\begin{equation}
Q \sim 4\times10^{-3}\frac{\kappa_{\rm ring}}{\sigma_{\rm ring}}
\left(\frac{C_{\rm s}}{\mbox{10 km s$^{-1}$}}\right),
\end{equation}
where $\sigma_{\rm ring}$ is the surface density of the gas ring and
$\kappa_{\rm ring}$ is the epicycle frequency at the radius of the gas ring.
The radius of the gas ring is about 0.03 in model B, and $\kappa (R=0.03)=145$.
Therefore, if $\sigma_{\rm ring} \ga 0.6$,
$Q$ becomes less than one, and the gas ring is expected to be gravitationally unstable.
Since the surface density of the gas ring is about 100 times larger than
of the initial gas disk, 
gravitational instability will occur in the ring if the surface density
of the initial disk $\sigma_{\rm init}$ is greater than 0.006.

In this paper, we adopt two values for initial surface density, i.e.
a supercritical value $\sigma_{\rm init}=0.01$ and a subcritical value
$\sigma_{\rm init}=0.0035$.
In physical units, $\sigma_{\rm init}=12 M_\odot{\rm pc^{-2}}$ for the
supercritical case, and $\sigma_{\rm init}=4.2 M_\odot{\rm pc^{-2}}$ for
the subcritical case.
The mass of one SPH-particle is $0.9\times10^4 M_\odot$ for the
supercritical surface density case.

\section{Results}

\subsection{Evolution of the gas disk}

The evolution of the outer part of the gas disk ($R \ga 0.1$) for the two cases
is almost the same as that of the non-self-gravitating model in Paper I.
This is because the self-gravity of the gas is too weak for the outer part
of the gas disk to affect the gas dynamics.
A snapshot of the whole gas disk for the supercritical case at $t=0.8$ is
shown in Fig. 1.
The straight shocks parallel to the bar ($R<0.3$) are seen at the leading edge of
the stellar bar.
This is the same result as in Paper I.
Two leading spiral shocks seen at $R>0.3$ are also seen in the
non-self-gravitating models of Wada \& Habe (1992).
These leading spiral shocks are not seen in Paper I, because we treated a
smaller gas disk ($R=0.3$) in Paper I.

\subsubsection{The subcritical case}

The time evolution of the central region ($R \la 0.1$) of the gas disk for the
subcritical case is shown in Figs. 2.
In the case of subcritical surface density, the evolution
of the gas disk is as follows.
At $t=0.3$ two straight shocks excited by the NLR are already formed.
Then gas falls along these shocks, and form a gas ring that has the radius
of $R=0.03$.
Although the gas ring has a weakly perturbed structure,
the formation process of the gas ring is almost the same as that in the
non-self-gravitating model in Paper I.

\subsubsection{The supercritical case}

The nuclear gas ring becomes unstable in 
the supercritical model ($\sigma_{\rm init}=0.01$) as we expected (Fig. 3).
In the early stages ($t < 0.4$),
the evolution of the gas disk is almost the same as that of the 
non-self-gravitating model with the same parameters (model B in Paper I),
since the self-gravity of the gas disk is not strong at this time.
From $t=0.3$, a gas ring begins to form at a radius $R=0.03$ around the SMBH.
At $t=0.4$, gas clumps are already formed in two dense regions where the gas
ring contacts with the straight shocks due to the NLR.
The clumps move along the ring.
Some diffuse clumps are destroyed by the tidal force of the SMBH,
and the debris is elongated almost along the gas ring orbit.
Other dense and compact gas clumps increase their mass via mass accretion of
surrounding gas and collisions with other clumps.
Several clumps merge into a massive gas clump at $t=0.6$.
The massive gas clump (hereafter we refer this clump as ``clump A'')
is at the position of $(x, y)=(-0.01, -0.025)$ for
$t=0.7$, and $(x, y)=(0.03, -0.005)$ for $t=0.8$ in Fig. 3.

The gas in the ring loses angular momentum by the torque from the clump A
and falls into the center.
By this process, a gas disk with very small radius of $R\sim0.01$ is formed
during $t=0.6\sim0.8$.
The nuclear gas disk has weak spiral structure (see also Fig. 7(b)),
and the mass of the disk is of the same order as the massive gas clump at the
final time of the simulation.
During the growth of the clump A, 
it also gains specific angular momentum.
This is obvious in Fig. 4(a),
in which we show the time evolution of the specific angular momentum versus
the mass of the clump A.
We find that the mean distance of the massive clump from the center
slightly increase as it gains angular momentum.
Fig. 4(b) shows the time evolution of the angular momentum of gas in
$R<0.035$ at $t=1.0$ including the massive gas clump.
The decrease of the total angular momentum of the gas is due to the torque from
the bar potential. It is clear that the clump A gains angular momentum from
the rest of the gas inside $R=0.035$.
This process dominates
the exchange of angular momentum between the gas and the stellar bar.
That is, the infall of the gas and the 
formation of the nuclear gas disk ($R<0.01$), which is the final
phase of the bar-driven fueling, are dominated by self-gravity of the gas.

After $t=0.8$, the nuclear gas disk evolves gradually, but the size of the 
nuclear disk does not change drastically.
Therefore, we stop the simulation at $t=1.0$.

Fig. 5(a) and 5(b) show the time evolution of the gas mass within $R=0.035$
and $R=0.01$ in the two cases of our simulations and in the non-self-gravitating
case.
The value $R=0.035$ and $R=0.01$ correspond to the radii of the gas ring and
the nuclear gas disk in the supercritical case.
The gas mass is normalized to the value at $t=0$.
Fig. 5(a) shows that for the nuclear ring there is no significant difference between the three
models.
In other words, the formation process of the gas ring is not affected by
the self-gravity of the gas.
In Fig. 5(b), the normalized gas mass for the subcritical case and the
non-self-gravitating case remain small,
since in both cases the gas does not concentrate into the region with $R<0.01$.
The evolution of the supercritical case, on the other hand,
is similar to above only until the time $t=0.6$.
However, the mass in the nuclear region ($R <0.01$) 
increases abruptly from $t=0.6$ to $t=0.75$.
The time of this abrupt increase corresponds to the formation epoch of the
nuclear gas disk.
During the same time, the clump A also forms in a similarly 
abrupt way (Fig. 4(a)).
This correspondence supports the interpretation that the nuclear gas disk is
formed by the influence of the clump A.
The mass of the nuclear gas disk at $t=1.0$ is $4.2\times10^7 M_\odot$.
The average accretion rate to the nuclear region ($R<50$ pc) 
in this stage is about $1 M_\odot$ yr$^{-1}$ for $3.5\times 10^7$ yr.

As mentioned above, the redistribution of the gaseous angular momentum 
in the nuclear region is crucial for the formation of the nuclear disk.
In order to confirm this, we removed the clump A at $t=0.6$,
and continued the calculation.
The evolution of the gas mass inside $R=0.01$ in this case is shown
in a dash-dot curve in Fig.5(b).
The gas infall slows down from $t=0.6$ to $t=0.8$,
but it recovers after $t=0.8$.
This is because the infalling gas from the outside of this region forms a new
gas clump, and it works for redistribution of the angular momentum after
$t=0.8$, when the mass of the new gas clump attains the same order of that of
the massive gas clump removed at $t=0.6$.

\subsection{Kinematic structure of the nuclear gas}

Figs. 6 and 7 are the position-velocity diagram and iso-velocity contour for
the two cases at $t=1.0$, respectively.
In the position-velocity diagram for the subcritical case (Fig. 6(a)), most of
gas is concentrated in the central linear structure which connects two circular
velocity curves. This linear structure means
axisymmetric rotation of the gas ring around the SMBH.
The weak structure seen at the retrograde velocity region comes from the 
straight shocks,
along which the gas has strong non-circular motion due to the NLR.
The iso-velocity contours (Fig. 7(a)) of the gas ring is almost symmetric.
This also shows that the gas ring rotates almost circularly.
The weak spiral waves in the gas ring can be seen as `spikes' in the
iso-velocity contours.
These two diagrams show that the self-gravity of the gas ring is not 
strong enough to destabilize the ring structure in the model with the
subcritical surface density.

In the supercritical case, on the other hand,
there are many linear structures which are almost
vertical in the position-velocity diagram (Fig. 6(b)).
These linear structures correspond to gas clumps with rapid spin,
which are supported by the spin against their self-gravity.
Most clumps including the clump A have prograde spin
as naturally expected from fragmentation of the rotating gas disk.
Iso-velocity contours of this model in Fig. 7(b) show that the nuclear gas disk
rotates almost circularly around the SMBH.
The irregular structure in the isovelocity contour seen in the outside of the
nuclear gas disk ($R>0.02$) is a numerical artifact, because there is only a
small number of SPH particles at this radius.
However, spikes associated with the spiral waves can be seen.
The strong spin of the clump A causes strong distortion of the
contour at $(x,y) = (-0.015,-0.03)$.

\section{Discussion}

\subsection{Summary}

We have made numerical simulations of the gas disk considering the self-gravity
of gas in oder to investigate the further evolution of the nuclear gas ring
formed by the effect of the NLR.

For an initial gas disk with subcritical surface density, the gas ring behaves
in almost the same way as in the non-self-gravitating model described in
Paper I.
Although the gas ring shows weakly clumped structure, the self-gravity of gas is
not efficient enough to change the gas ring structure drastically.
On the other hand, in the model with the supercritical surface density,
clumps are formed in the gas ring, and gas fueling to the galactic
center occurs due to the torque from the massive gas clump.
The massive clump which occupies a large fraction of the gas mass in the nuclear
region is formed by merging of gas clumps and accretion of surrounding gas.
We find that the redistribution of angular momentum of the gas in
the nuclear region is a key mechanism of the fueling of the SMBH, and 
self-gravity of the gas play a crucial role in this process.
The massive gas clump remains near the radius of the gas ring, and shows no sign
of falling into the center in our simulation.
The small nuclear gas disk is formed as a result of the gas fueling.
The evolution of the subcritical and supercritical surface density cases
are reasonably understood if the critical surface density 
is the same as that is deduced from the
Toomre's $Q$ value using the results of non-self-gravitating models.
This suggests that the Toomre's $Q$ value is useful to judge the 
stability of the nuclear gas ring in a weak bar, although
$Q$ is derived from the linear analysis of an axisymmetric mode of
gravitational instability.

Finally we should emphasize that the central mass concentration is not 
necessarily a super massive black hole.
A massive star cluster or a dense molecular cloud can also produce the NLR,
and affect the gas dynamics similarly.

\subsection{Smaller SMBH case}

The critical surface density in the case of a smaller SMBH mass than 
assumed in this paper ($M_{\rm BH} = 0.01 M_{\rm axi})$ would be
an interesting problem.
We have found that the actual critical surface density
of the gas ring for the gravitational instability 
is consistent with one estimated by using Toomre's $Q$ value of
the gas ring in the non-self-gravitating model.
Assuming that this is also the case in models with a smaller SMBH,
we can estimate the critical surface density in the smaller
SMBH case.
In Paper I, we have calculated the case of smaller SMBH
 ($M_{\rm BH}= 0.001 M_{\rm axi})$  without self-gravity of gas (model Ba).
The radius of the gas ring formed in model Ba is about $0.01$.
The epicycle frequency $\kappa$ at this gas ring is $229$, and the surface
density is about 100 times of the initial value.
Therefore, the condition of this gas ring to be gravitationally unstable for 
$M_{\rm BH} = 0.001 M_{\rm axi}$ is,
using equation (5), $\sigma_{\rm init} \ga 0.01$.
This is about factor two larger than in the 
model presented here
 (i.e. the critical surface density for the initial gas disk equals to $0.006$ for $M_{\rm BH} = 0.01 M_{\rm axi}$). 
In other words, the supercritical case in this paper ($\sigma_{\rm init}=0.01$)
is marginally stable, if the SMBH is factor 10 smaller.
Therefore, we expect both an unstable gas ring followed by the fueling for the 
center and a stable or marginally stable nuclear ring, depending on the mass of SMBH and
the initial gaseous density. Observational implications about this are
discussed in section 4.4.

\subsection{Effects of dynamical friction on the massive gas clump}

We do not consider dynamical friction of the gas clumps 
due to stellar components in our simulations,
since we do treat the stellar component as a static potential.
Dynamical friction may affect the fate of the massive gas clump.
Stark et al. (1991) discussed the dynamical friction time scale of
massive molecular clouds in a central kpc region of our Galaxy.
Applying their estimate to our case, 
the time scale of the massive gas clump in our simulation to spiral into
the center by dynamical friction is
\begin{equation}
t_{\rm fric}=6.7\times10^7 \left(\frac{{\rm ln} \Lambda}{6}\right)^{-1}\left(\frac{m_{\rm c}}{10^7 M_\odot}\right)^{-1}
\left(\frac{v_{\rm c}}{100{\rm km\ sec^{-1}}}\right)^3\left(\frac{\rho}{10 M_\odot {\rm pc^{-3}}}\right)^{-1}{\rm yr},
\end{equation}
where ${\rm ln} \Lambda$ is the usual Coulomb logarithm.
$m_{\rm c}$ and $v_{\rm c}$ are the mass and the circular velocity of the
massive gas clump, respectively.
$\rho$ is the stellar mass density.
At $t=0.6$, $m_{\rm c}=7.4\times10^6 M_\odot$ (Fig. 4(a)),
$v_{\rm c}=160{\rm km\ sec^{-1}}$ and thus $t_{\rm fric}=2.7\times10^8$ yr.
Since this time scale of dynamical friction is larger than the orbital period
of the massive gas clump $t_{\rm orb}=3.8\times10^6$ yr,
the orbital motion of the massive gas clump is not changed by
dynamical friction within a few orbital period.
This time scale is also larger than the period of the rapid increase of gas
mass within $R=0.01$ ($t=0.6-0.75$, Fig. 5(b)) due to the torque from the
massive gas clump, $t_{\rm fuel}=2.6\times10^7{\rm yr}$.
Therefore, the rapid gas fueling process does not seem to be affected by the
dynamical friction.

However, dynamical friction can be effective after the rapid gas fueling
because $t_{\rm fric}$ becomes shorter as the mass of the massive gas clump
increases.
At $t=1.0$, since $m_{\rm c}=4.2\times10^7 M_\odot$ and $v_{\rm c}=
114 \; {\rm km\ sec^{-1}}$,
$t_{\rm fric}=2.4 \times 10^7$ yr.
The orbital period of the massive gas clump at $t=1.0$ is
$9.6\times10^6$ yr,
so that the massive gas clump is expected to fall into the center in 
about 2-3 orbital periods after $t=1.0$.
This infall can cause further gas fueling to the central engine,
and the AGN can be activated.
We will study this process in the subsequent paper.

\subsection{Implications to observations}

In early stages ($t\sim0.4$) of both the subcritical and the supercritical
surface density cases, two dense gas regions are formed in the ring near the
bar minor axis because of the collision between the straight shocks and 
the gas ring.
This situation seems to correspond to the `twin peaks' structure (Kenney et al.
1992).
We could argue that `twin peaks' are found in early stage of gas fueling
process in weakly barred galaxies with SMBHs.

In the subcritical model, we find the formation of the nuclear gas ring and
two straight shocks parallel to the bar.
This situation is similar to the gaseous and star-forming structure in
IC342 (Ishizuki et al. 1990; B\"oker, F\"orster-Schreiber and Genzel 1997).
Ishizuki et al. (1990) showed that IC342 has a pair of narrow ridges of
molecular gas and a ring-like structure ($R \sim 70 {\rm pc}$) in the nuclear
region.
Intense star formation takes place in the ring.
B\"oker, F\"orster-Schreiber and Genzel (1997)
revealed that the IR images of the central star cluster of IC342 show a steep
concentration at the center, i.e. central mass concentration, and a bar-like
elongation with an axis ratio of about 1.25 on a scale of at least 110 pc.
They also estimated the dynamical mass within the radius of 40 pc to be
$1.1\times10^6 M_\odot$.
The radius of the star-forming ring is at least a factor of 5 smaller than
that of other typical star formation rings which are thought to be formed by
the effect of the ILRs.
Therefore, the nuclear structure of IC342, i.e. the pair of narrow ridges and
the ring, seems to correspond to the gas ring formed in the subcritical case
with smaller SMBH discussed in section 4.2.

Some authors have assumed that the radius of observed ring-like structures
corresponds to the location of some resonances (e.g. van den Bosch \& Emsellem
1998).
This is not the case for the NLR, because the gaseous ring that is formed by
the effect of the NLR has about 5 times smaller radius than that of the NLR,
as shown in Paper I. 
In general, axisymmetric ring orbits of the gas at Lindblad resonances are
unstable, because the perturbation from the bar becomes maximum there.
Deviation from the circular orbit becomes maximum at 
the resonances (Wada 1994). 
An elongated ring can be formed near the inner ILR, but this is not the
case for the outer ILR. Typical response near the outer ILR is
trailing spirals.
We suggest, therefore, that the observed axisymmetric ring-like 
starforming region does not mean a resonance at the ring radius.

Recent Hubble Space Telescope (HST) observations with high angular resolution
revealed fine structures in nuclear starburst regions of barred galaxies, e.g.,
NGC 4303 (Colina et al. 1997), NGC 1365 (Kristen et al. 1997) and
NGC4314 (Benedict et al. 1993, 1998).
NGC 4303 has a nuclear one-armed spiral structure (Colina et al. 1997).
In our model of the supercritical surface density,
the gas which loses angular momentum by the torque from the massive gas clump
is elongated by the tidal force of the SMBH, and a one-armed structure
is formed.
The time scale of the one-armed structure is almost same as its rotation period
$t\sim 10^6$ yr.
If star formation occurs in the gas,
a nuclear one-armed star forming region like in NGC 4303 could be formed.
If this is the case, 
a massive clump-like structure of gas (e.g. a giant molecular cloud)
should be observed in the nuclear region of NGC 4303.
Another possible scenario to produce the one-arm star forming region is
collisions between the massive clump and the nuclear disk.
Due to the dynamical friction from stars, the massive clump is expected to 
sink toward the center (see section 4.3).
Eventually the clump will collide with the nuclear disk, and star formation
due to the collision would be expected. 
This collision may cause one-armed star forming regions as similar as
that caused by a collision between a massive gas disk around SMBH and
an infalling secondary SMBH (Taniguchi, Wada 1996).

The mass of the most massive gas clump is $4.2 \times 10^7 M_\odot$ in our
simulation.
We showed that the massive gas clump collides with other smaller clumps.
Collision between massive clumps may induce active star formation.
Therefore, `super star clusters' (SSCs) observed in starburst galaxies
(Maoz et al. 1996 and references there in) may be
formed from the massive gas clump, if active star formation occurs in the
collision process.
Kristen et al. (1997) found SSCs in the nuclear region of NGC 1365.
NGC 1365 is a barred spiral galaxy with an active nucleus (i.e. SMBH),
and the SSCs are observed in the nuclear ring around the nucleus.
This situation resembles our results.
We suspect that the SSCs in NGC 1365 could be formed in the process of the
bar-driven fueling and subsequent gravitational instability.

\section*{Acknowledgments}

We thank T. B\"oker for fruitful discussions and suggestions in the
original version of this paper.
We also thank C. Norman \& {\it the HOT TOPICS} group for stimulating discussions.
Numerical computation in this work is carried out at the Yukawa Institute
Computer Facility and the parallel computer in the Division of Physics,
Graduate School of Science, Hokkaido University.
This work is supported in part by Research Fellowships of the Japan Society for
the Promotion of Science for Young Scientists (No. 1318), and the Grant-in-Aid
for Scientific Research of the Japanese Ministry of Education, Science, Sports
and Culture (No. 07640344, 08874014).
KW thanks Yamada Science Foundation for the support while at STScI.

\clearpage

\begin{figure}

\caption{A snapshot of the distribution of the gas at $t=0.8$ in the
supercritical surface density model ($\sigma_{\rm init} = 0.01$).
The bar major axis is horizontal.
The radius of the NLR is $0.14$, and the inner ILR is located at $R=0.37$.
Unit of the length and time are 5 kpc and $1.75\times10^8\ {\rm yr}$.}

\caption{Time evolution of the gas distribution in the central 1 kpc $\times$
1 kpc region of the subcritical surface density model ($\sigma_{\rm init}=
0.0035$) in the rotating bar coordinates.
The bar major axis is fixed horizontally on each frame.
Time is shown  at the upper left of each figure.
The unit of time is $1.75\times10^8\ {\rm yr}$.
The radius of the NLR is $0.14$.}

\caption{Same as Fig. 2, but for the supercritical surface density case ($\sigma_{\rm init} = 0.01$).}

\caption{
(a) Time evolution of the specific angular momentum and the mass of the
massive gas clump ('clump A').
(b) Time evolution of the angular momentum of gas that is in $R<0.035$ at
$t=1.0$.
The lines with crosses, stars and diamonds show the angular momentum of the
total gas, the clump A and the gas other than the clump A,
respectively.
}

\caption{
Time variation of the gas mass in (a) $R<0.035$ and (b) $R<0.01$ for the
non-self-gravitating model, the subcritical ($\sigma_{\rm init}=0.0035$) and
the supercritical surface density model ($\sigma_{\rm init}=0.01$).
The gas mass in each model is normalized by their initial values.
The unit mass for the supercritical surface density case is
$4.1\times10^{-4}$ in (a), and $3.3\times10^{-5}$ in (b).
The dash-dot line in (b) shows a test case,
 in which we remove the most massive gas clump at $t=0.6$.
See text for detail of the test.}

\caption{
(a) Position velocity diagram ($x$ vs. $v_y$) of the subcritical surface
density case at $t=1.0$.
The $x$-axis is the bar major axis, and the $y$-axis is the bar minor axis.
The SPH particles within $R=0.1$ are plotted.
The solid curve shows the circular velocity.
(b) Same as (a), but for the supercritical surface density case.}

\caption{
(a) The iso-velocity contours of $v_y$ for the subcritical surface density case
superposed on the gray-scaled surface density distribution of gas at $t=1.0$.
The contour levels are from $-10$ to $10$ in our units.
(b) Same as (a), but for the supercritical surface density case.
The massive gas clump is at (-0.013, -0.031).}

\end{figure}

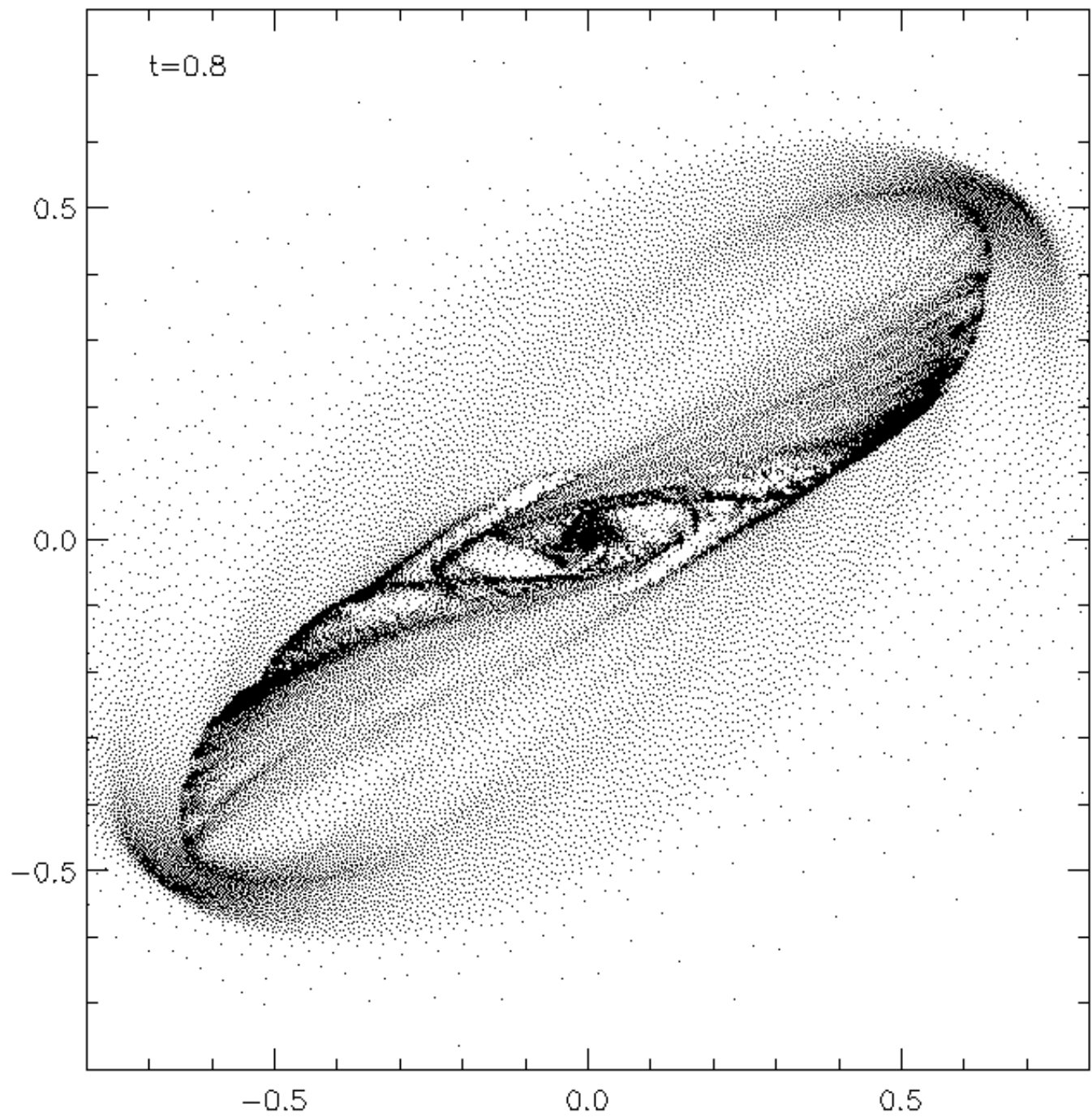
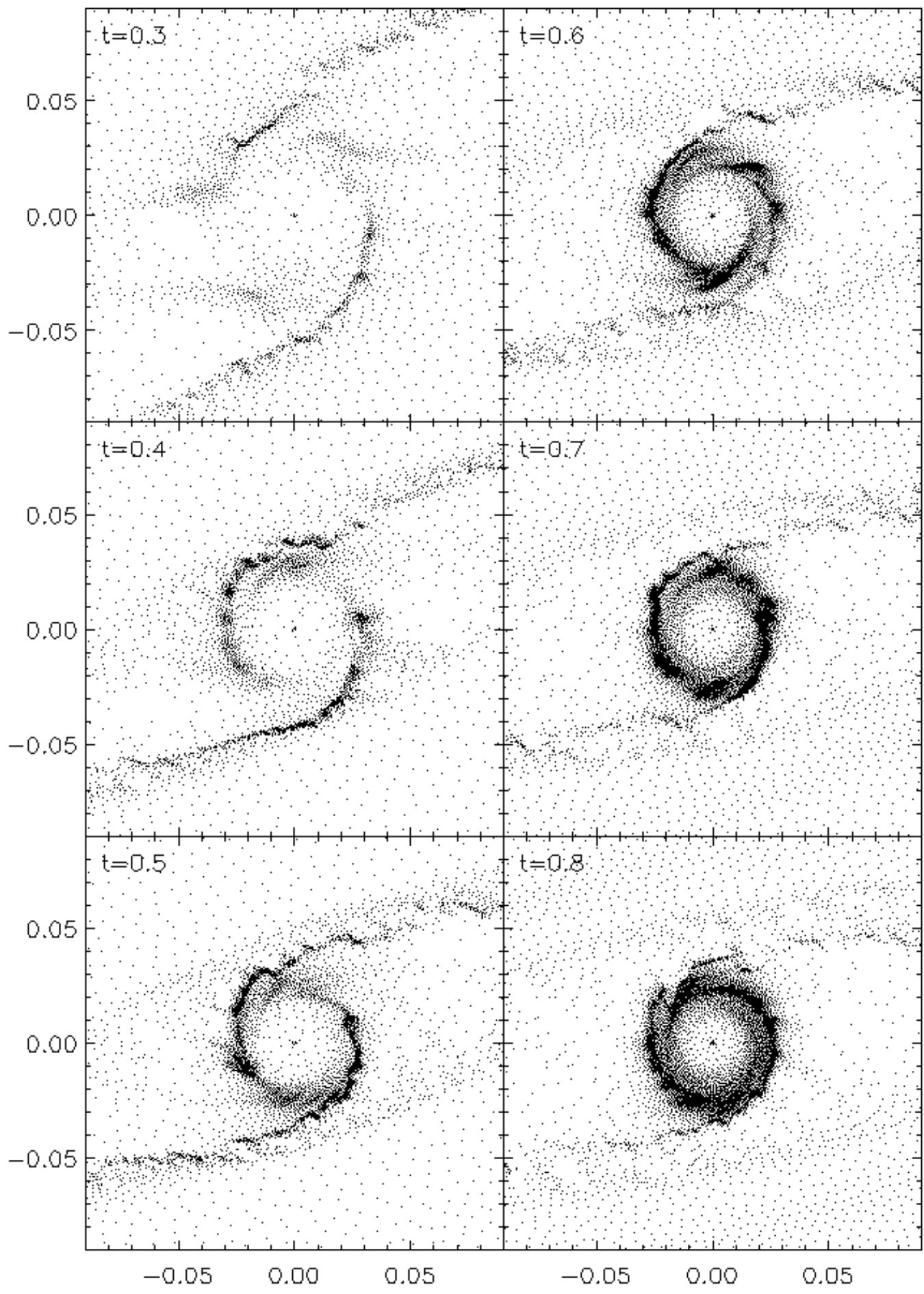
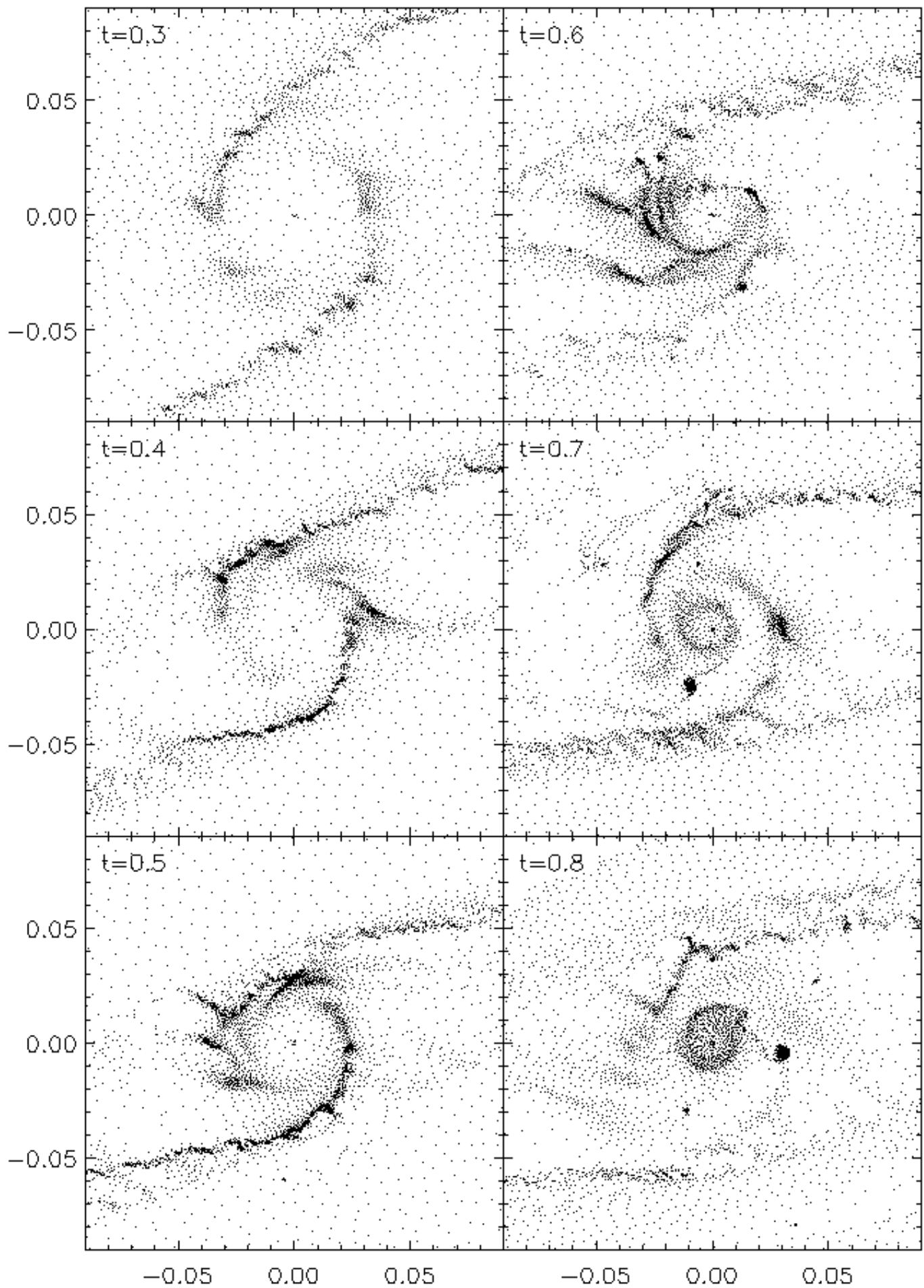
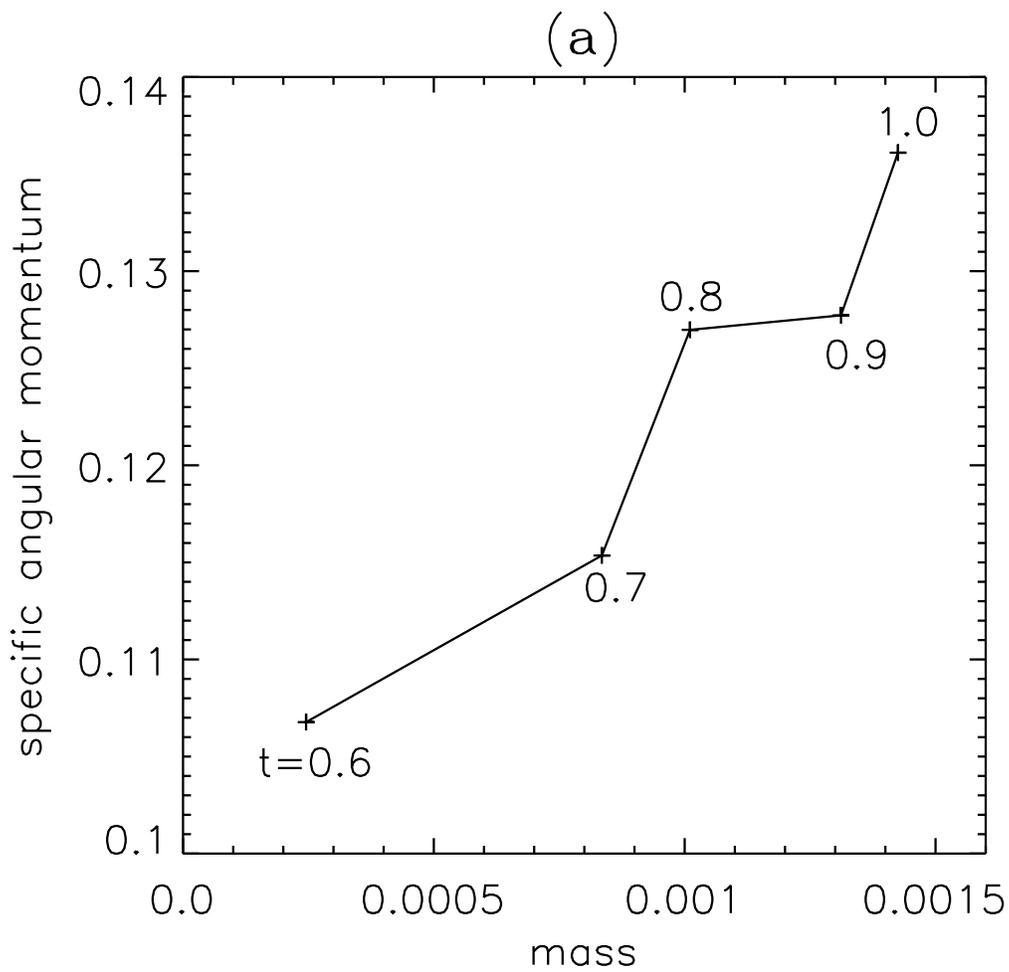
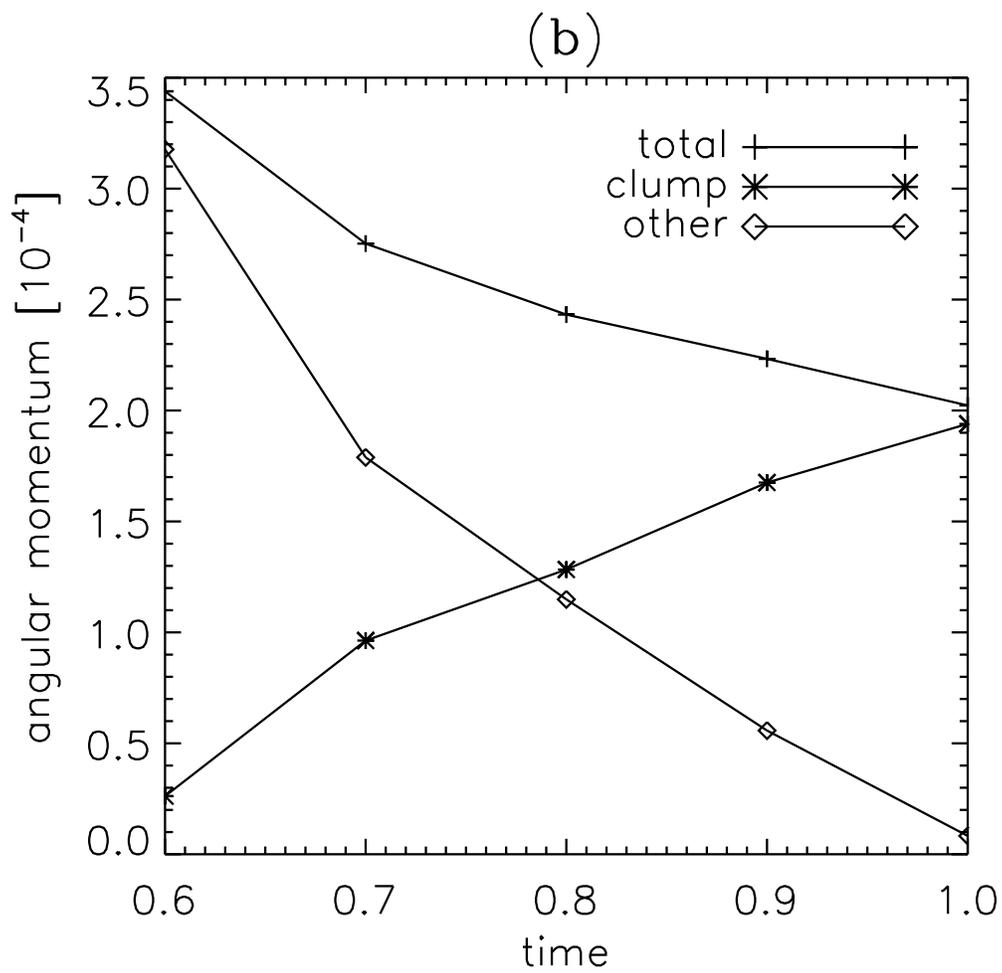
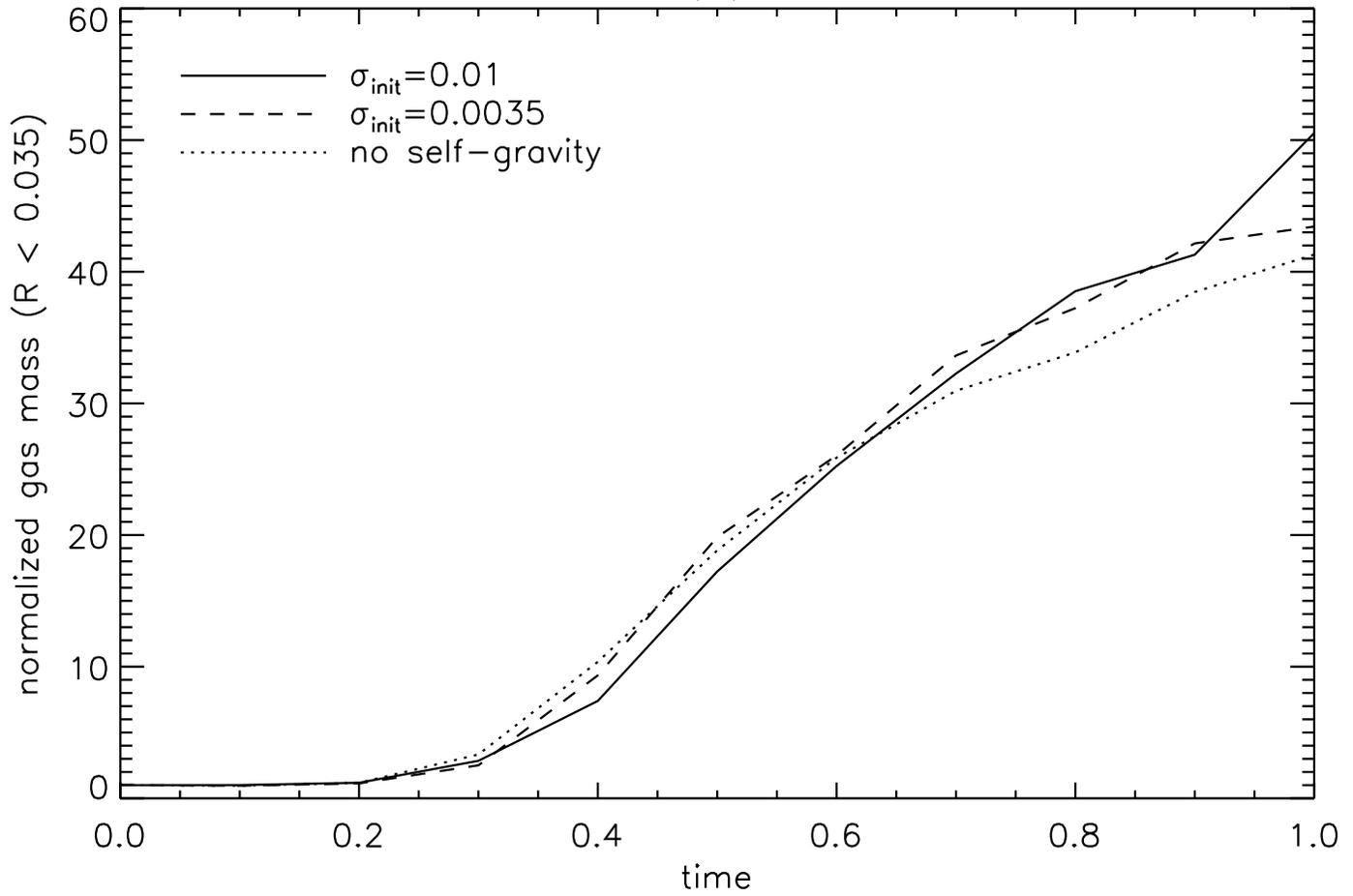
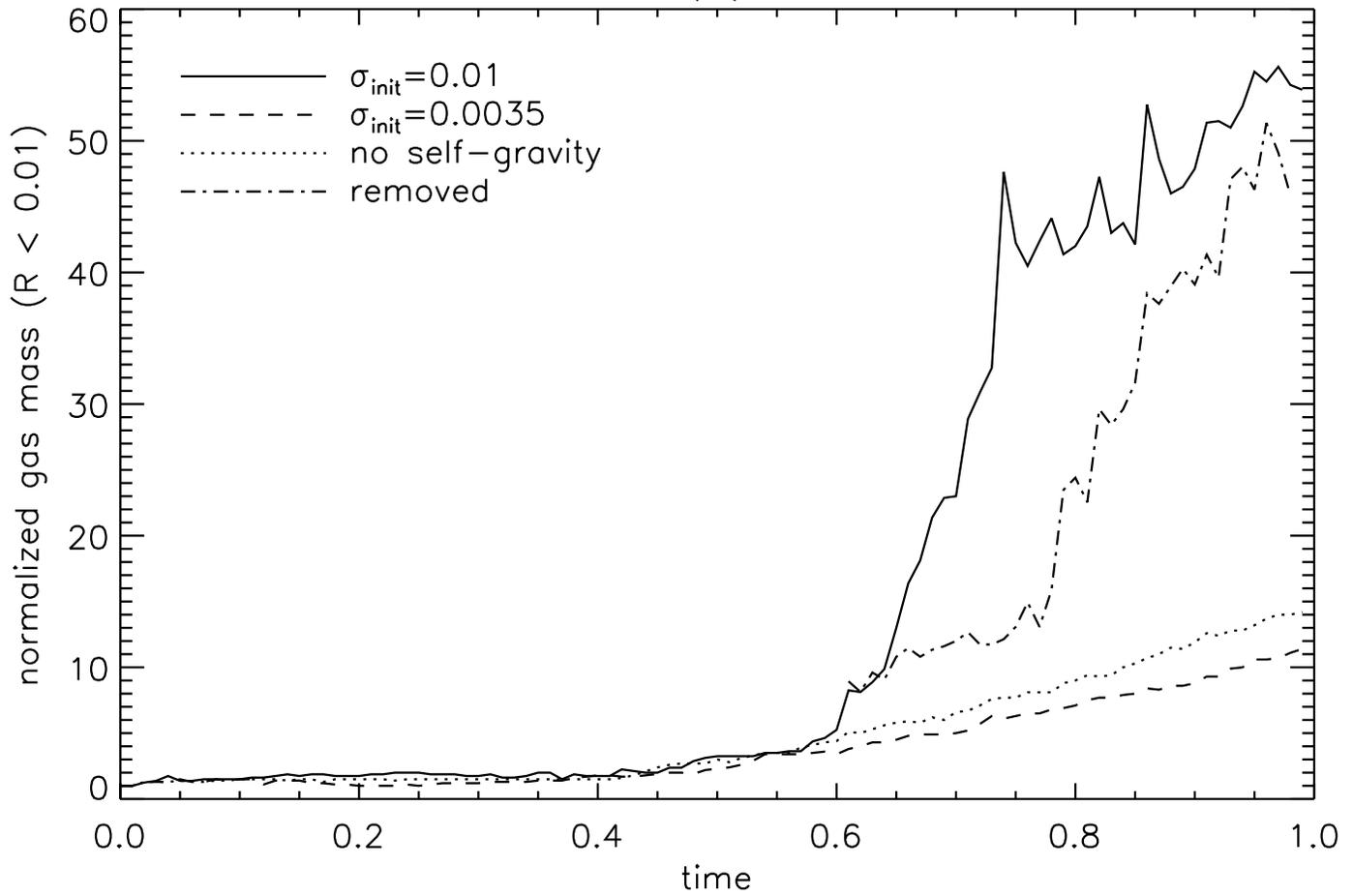
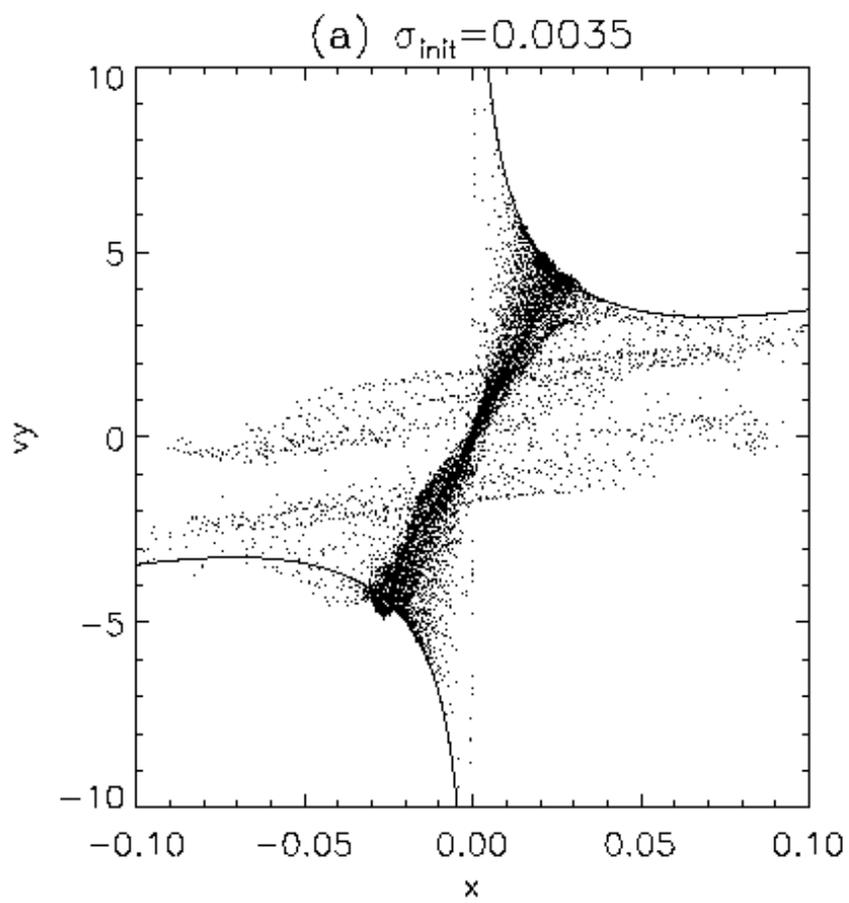
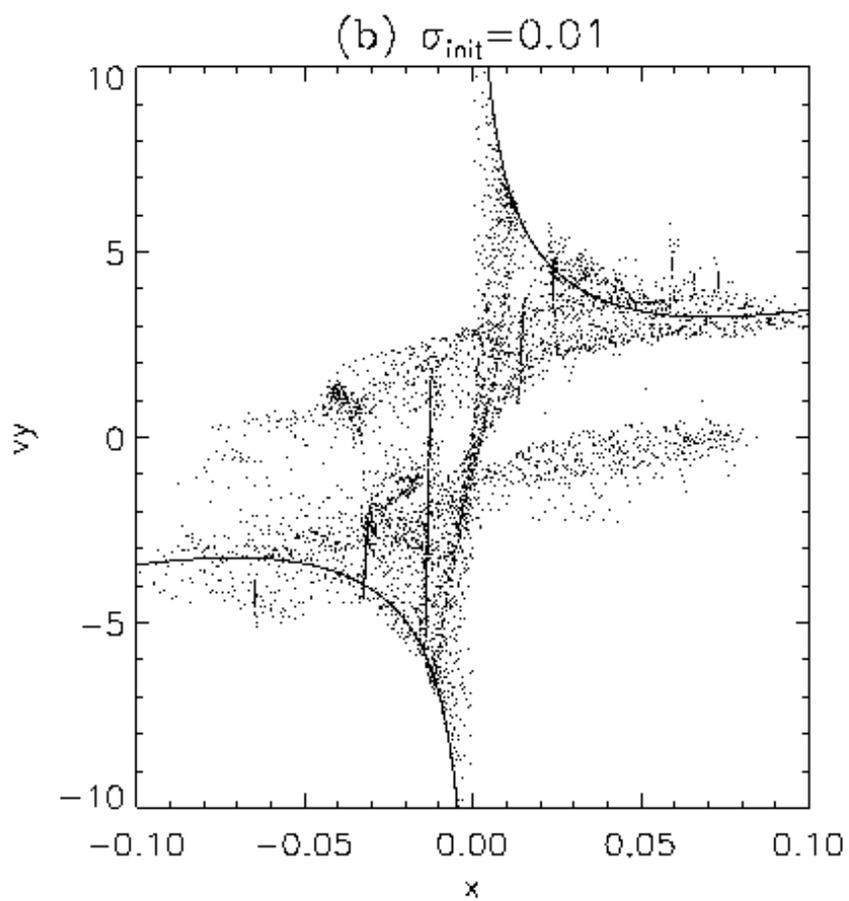
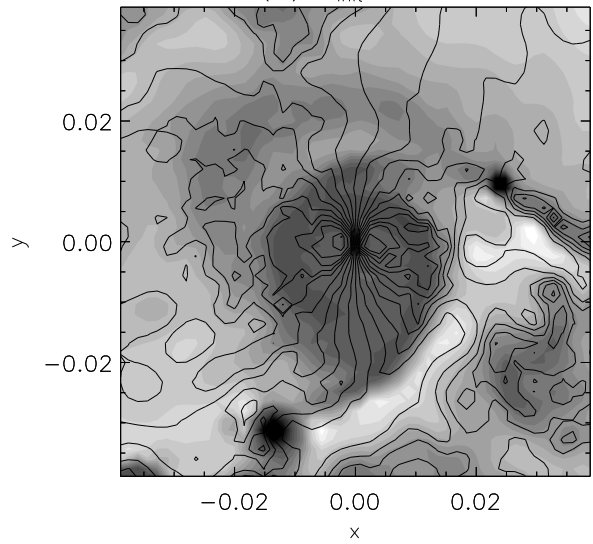
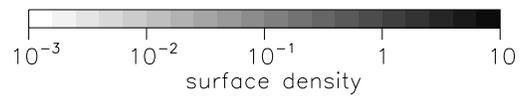
\end{document}